\documentclass[11pt,a4paper]{article}
\usepackage{amsmath}
\usepackage{comment}
\usepackage{graphicx}
\usepackage{cite}
\usepackage{indentfirst}
\usepackage{xcolor}
\usepackage[margin=0.8in]{geometry}
\usepackage[colorlinks,linkcolor=blue,anchorcolor=blue,citecolor=blue,urlcolor=blue]{hyperref}
\usepackage{bm}
\begin{document}
\section*{Topological magnons in a non-coplanar magnetic order on the triangular lattice}
\vspace{\baselineskip}
 Linli Bai$^{1,2}$, Ken Chen$^{1,2,*}$
\vspace{\baselineskip}

$^1$ School of Physical Science and Technology $\&$ Key Laboratory for Magnetism and
Magnetic Materials of the MoE, Lanzhou University, Lanzhou 730000, China

$^2$ Lanzhou Center for Theoretical Physics, Key Laboratory of Theoretical Physics of Gansu Province,
$\&$ Key Laboratory of Quantum Theory and Applications of MoE,
Lanzhou University, Lanzhou, Gansu 730000, China

*Corresponding author E-mail: chenk20@lzu.edu.cn
\subsection*{Abstract}
The bond-dependent Kitaev interaction $K$ is familiar in the effective spin model of transition metal compounds with octahedral ligands.
In this work, we find a peculiar non-coplanar magnetic order can be formed with the help of $K$ and next-nearest neighbor Heisenberg coupling $J_2$ on the triangular lattice.
It can be seen as a miniature version of skyrmion crystal, since it has nine spins and an integer topological number in a magnetic unit cell.  
The magnon excitations in such an order are studied by the linear spin-wave theory. 
Of note is that the change in the relative size of $J_2$ and $K$ produces topological magnon phase transitions although the topological number remains unchanged.
We also calculated the experimentally observable thermal Hall conductivity, and found that the signs of thermal Hall conductivity will change with topological phase transitions or temperature changes in certain regions.

\vspace{\baselineskip}

Keywords: non-coplanar magnetic order; topological magnons; thermal Hall effect.

\section{Introduction}
Magnon is the quantum of low-energy collective excitation in a magnetic system\cite{Bloch1930,Kubo1952,Dyson1956}.
Since a magnon does not generate Joule heating during motion and has a much longer diffusion length than an electron\cite{Kruglyak2010,Chumak2015,PengYan2021}, it has application prospects for storing and disseminating information in the future \cite{PengYan2021}.
Inspired by topological insulators which can support chiral edge/surface states that are immune to backscattering\cite{Hasan2010,XLQi2011}, researchers are also committed to find corresponding phenomena in magnons\cite{Lifa2013,Mook2014,Chisnell2015,Nakata2017,Bo2018,McClarty2018,Cai2021,Gomez2021,McClarty2022,Koyama2023}.
Topologically non-trivial magnons can be driven by a thermal gradient, forming transverse heat currents by the Berry phase mechanism \cite{Matsumoto2011prl}. This is called the thermal Hall effect\cite{Katsura2010,Onose2010,Matsumoto2011,Ideue2012,Shindou2013,Matsumoto2014,Hirschberger2015,Hirschberger2015prl,Owerre2017,Murakami2017,Mook2019,Mook2022,Zhuo2023}, which has been experimentally observed in pyrochlore \cite{Onose2010,Ideue2012} and kagome \cite{Chisnell2015,Hirschberger2015prl} ferromagnets. Relevant research has been in full swing\cite{Murakami2017,zhang2023thermal}.

The non-planar spin textures with non-zero topological numbers are also a key research focus in condensed matter physics\cite{Skyrme1962,NagaosaTokura2013,FertRC2017,Zhou2018,BogdanovPanagopoulos2020,GobelMT2021,YU20231}, and the most representative one is the skyrmion\cite{Skyrme1962,NagaosaTokura2013}.
In the continuous limit, a skyrmion is topologically protected, which means it cannot be generated or removed by any continuous deformation\cite{NagaosaTokura2013}.
In the actual magnetic system, although the lattice is discrete and the size of the skyrmion is limited, a skyrmion is still relatively stable, which allows it to be manipulated independently as a quasi-particle\cite{Je2020}.
In recent years, researchers have begun to pay attention to the magnon excitations in the skyrmion \cite{PengYan2021}.
The topological magnons in ferromagnetic and anti-ferromagnetic skyrmion crystals have been discovered \cite{Molina2016,Daniel2019,Daniel2020,Asle2022,Timofeev2022prb,Akazaw2022,Weber2022}, and  the topological phase transitions caused by the interactions or magnetic field are discussed as well\cite{Daniel2020,Asle2022}. 
However, as far as we know, in previous works, researchers mainly involved the skyrmions in non-centrosymmetric magnets which are formed with the help of chiral Dzyaloshinskii-Moriya interaction.
This is just the tip of the iceberg of topological non-trivial spin textures\cite{GobelMT2021}. 
On the one hand, it has been proved that the skyrmion can also appear in the centrosymmetric magnets as a result of other interactions\cite{Kawamura2012,Utesov2021,Leonov2015,Hayami2016,Lin2018,Amoroso2020,Hayami2021,Wang2021,Hirschberger2021,Utesov2022,HAYAMI2022170036,hayami2023anisotropic}, such as dipolar interaction\cite{Kawamura2012,Utesov2021}, the single-ion anisotropy\cite{Leonov2015,Hayami2016,Lin2018,Amoroso2020,Hayami2021,Wang2021,Hirschberger2021,Utesov2022} or the bond-dependent interactions \cite{Yao2016,Amoroso2020}. 
On the other hand, beyond the skyrmion, there are many magnetic quasi-particles coming into the sight of researchers in recent years\cite{GobelMT2021,Rousochatzakis2016,Chen_2023}.
Whether topological magnons or relevant phase transitions exist within such spin textures remains to be further explored.

In this work, we study the interplay of Heisenberg and Kitaev interactions on the triangular lattice.    
The bond-dependent Kitaev interaction is widely considered in the theoretical model describing transition metal compounds with octahedral crystal field\cite{Winter_2017,TREBST20221,Razpopov2023}.
We find a peculiar non-coplanar magnetic order that can be formed by the competition between $K$ and next nearest neighbor Heisenberg coupling $J_2$.
There are nine spins in a magnetic unit cell and the topological number is $\pm 1$($\pm 2$), thus it can be seen as a miniature version of (high-Q) skyrmion crystal\cite{Aoyama2021}.
Then we focus on the magnon excitations in such order.
Through the linear spin-wave theory, we calculate the Chern number of each magnon band, and based on this, different regions are distinguished.
The change of $J_2$ will produce topological phase transitions although the topological number in real space remains unchanged.
Since thermal Hall conductivity is related to the magnon band topology \cite{Mook2022}, finally, we calculate the  thermal Hall conductivity and discussed its distinctions in different areas.
Thermal Hall conductivity is dominated by the Berry curvature in the lowest bands at low temperatures.
We also found that in certain regions, the signs of thermal Hall conductivity will change with topological phase transitions or temperature changes.
\begin{figure}[!h]	
	\begin{minipage}{1\linewidth}
		\centerline{\includegraphics[width=0.9\textwidth]{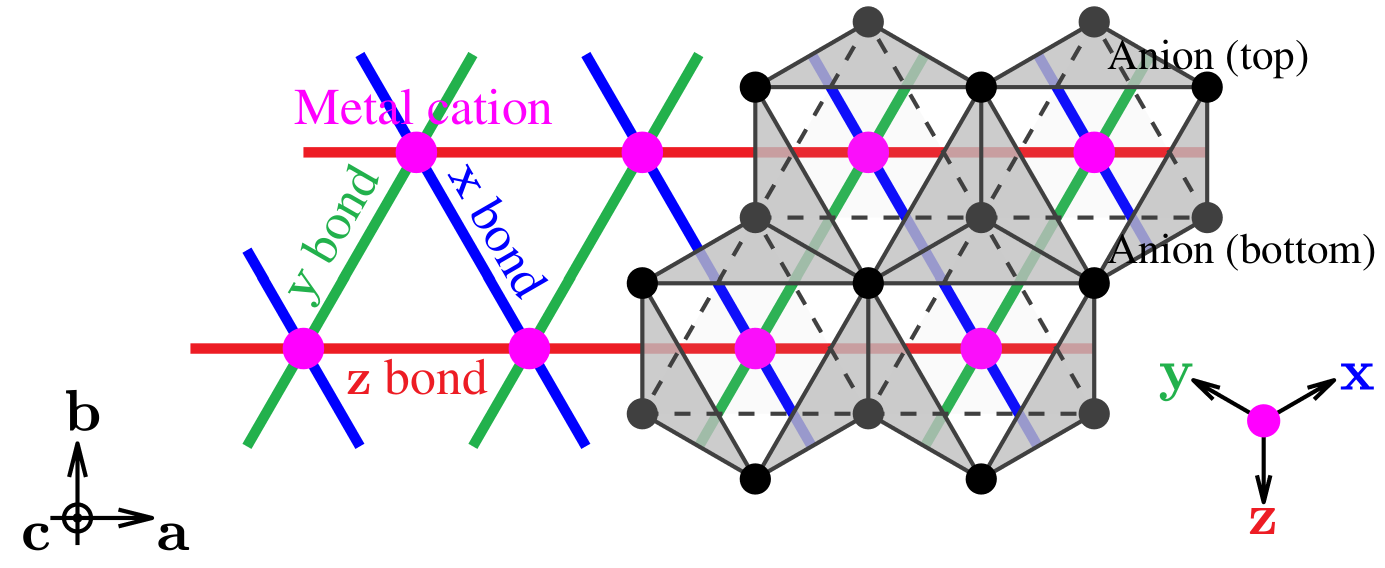}}
	\end{minipage}
	\caption{Schematic structure of the 
		Kitaev directions in a triangular lattice.
		The Kitaev interaction appears along with an octahedral crystal field where the metal cations are located at the center of octahedral ligands\cite{Razpopov2023,Kim2023}.
		The global coordinate system $\textbf{a}$-$\textbf{b}$-$\textbf{c}$ 
		is shown in the bottom left, where
		the triangular lattice composed of metal cations
		lies in the $\textbf{a}$-$\textbf{b}$ plane and $\mathbf{c}$ is perpendicular to  this plane.
		We mark the x, y, and z bonds in blue, green, and red, respectively.
In the $\textbf{a}$-$\textbf{b}$-$\textbf{c}$ coordinate  system, their corresponding Kitaev directions are given by $\textbf{x} = [{\sqrt2}/{2},\sqrt6/6,\sqrt3/3]$, $\textbf{y} = [{-\sqrt2}/{2},\sqrt6/6,\sqrt3/3]$, $\textbf{z} = [0,-\sqrt6/3,\sqrt3/3]$, respectively.}
	\label{fig1}
\end{figure}
\begin{figure}[!h]	
	\begin{minipage}{1\linewidth}
		\centerline{\includegraphics[width=0.95\textwidth]{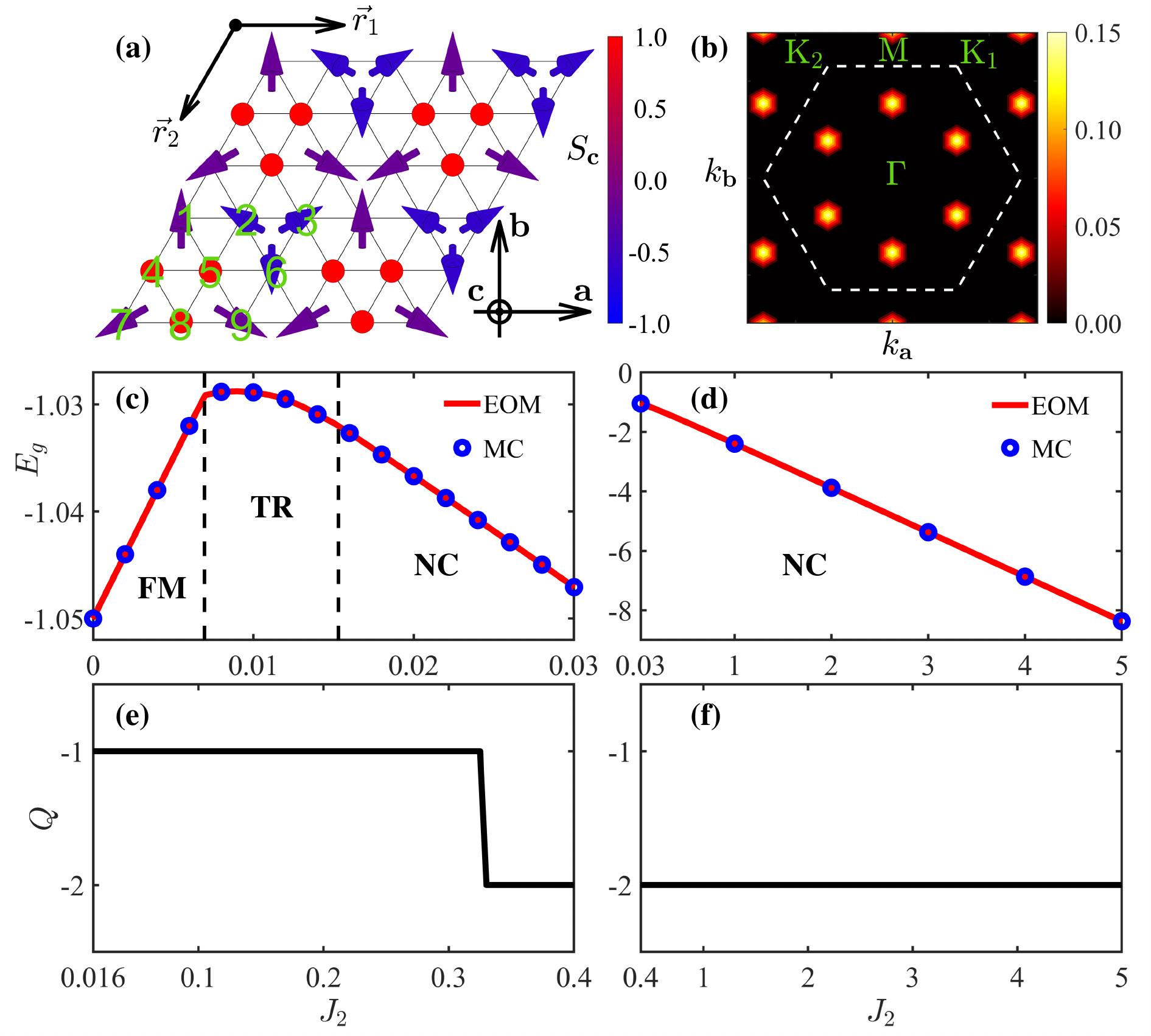}}
	\end{minipage}
	\caption{(a) Typical spin configuration of the non-coplanar order when $K=-1$, $J_2 = 1$, and $h=0.05$. The small arrows indicate
		directions of spins and their colors are based on the out of plane component. The magnetic unit cell includes nine spins.
		(b) The spin structure factor corresponds to (a), the peaks are located at the point of $2\mathbf{M}/3$.
		(c)-(d) The classic ground-state energy $E_g$ as a function of $J_2$ when $K = -1$, $h = 0.05$.  
		The result of energy optimization method (EOM) and Monte Carlo method is consistent.
		The ground state begins with the ferromagnetic (FM) order and rapidly transitions through a narrow transition region (TR) to the non-coplanar (NC) order.
		As $J_2$ increases further, the ground state remains in non-coplanar order.
		(e)-(f) The topological number $Q$ versus $J_2$ in the non-coplanar order region.  The topological number changes from $-1$ to $-2$ when $J_2 \approx 0.326$. }
	\label{fig2}
\end{figure}
\section{Model}
We consider the following model on a triangular lattice,
\begin{equation} \label{eq1:H}
\mathcal{H}= J_2\sum_{ \left \langle \left \langle ij \right \rangle \right \rangle} {\textbf{S}}_{i}\cdot {\textbf{S}}_{j}+K\sum_{{\left \langle ij \right \rangle}_\gamma}S_i^\gamma S_j^\gamma -\textbf{h}\cdot\sum_i \textbf{S}_i ,
\end{equation}
where $J_2$ is the exchange parameter of the next-nearest neighbor Heisenberg coupling. $\textbf{S}_i$ represents the spin at site $i$ and $S_i^\gamma=\textbf{S}_i\cdot\vec{\gamma}$, where
$\vec{\gamma} $ is the bond-dependent Kitaev direction.
For the three types of bonds x, y, and z, as shown in Fig.~\ref{fig1}, $\vec{\gamma} $ corresponds to $\textbf{x}$, $\textbf{y}$, and $\textbf{z}$, respectively.
The vectors $\textbf{x}$, $\textbf{y}$, and $\textbf{z}$ are perpendicular to each other, forming a coordinate system $\textbf{x}$-$\textbf{y}$-$\textbf{z}$.
$K$ is the exchange parameter of Kitaev interaction.
In practical materials related to triangular lattices, the Kitaev interaction appears along with octahedral ligands\cite{Razpopov2023,Kim2023}.
Accordingly, the Kitaev directions are fixed.
To better illustrate them, we introduce a global coordinate system
$\textbf{a}$-$\textbf{b}$-$\textbf{c}$.
In Fig.~\ref{fig1}, 
the $\textbf{a}$-$\textbf{b}$-$\textbf{c}$ and the $\textbf{x}$-$\textbf{y}$-$\textbf{z}$ coordinate systems are shown on
the bottom left and bottom right, respectively.
The relationship between the $\textbf{x}$-$\textbf{y}$-$\textbf{z}$ and the $\textbf{a}$-$\textbf{b}$-$\textbf{c}$ coordinates  systems satisfies
\begin{equation} 
\left( \begin{array}{c}  \textbf{x}\\\textbf{y}\\\textbf{z}\end{array}\right) =\left( \begin{array}{ccc}	\frac{\sqrt2}{2}  &  \frac{\sqrt6}{6}  &  \frac{\sqrt3}{3} \\ -\frac{\sqrt2}{2}  &  \frac{\sqrt6}{6}  &  \frac{\sqrt3}{3}  \\ 0  & -\frac{\sqrt6}{3}   &  \frac{\sqrt3}{3} \end{array}\right) \left( \begin{array}{c}  \textbf{a}\\\textbf{b}\\\textbf{c}\end{array}\right).
\end{equation} 
The last term is the Zeeman term caused by a magnetic field $\textbf{h} = h\mathbf{c}$, which is perpendicular to  the triangular lattice plane. 

We use the parallel-tempering Monte Carlo simulations\cite{Hukushima1996,YMiyatake1986} to uncover the spin textures.
After Monte Carlo simulations, the classic ground state is then obtained by iteratively aligning the spins with their local fields \cite{Janssen2016}. 
Other numerical energy optimization methods are also used to check if the energy is at a minimum.
Then we use the ground-state energy as well as the spin structure factor 
$\mathcal{S}_{\textbf{k}} =\frac{1}{N_{total}^2} \left| \sum_{i} \mathbf{S}_i e^{-i\textbf{k}\cdot\textbf{R}_i } \right|^2 $ 
 \cite{Kawamura2012,Shimokawa2019} to distinguish different phases. In the following, all the spin configurations are plotted in the $\textbf{a}$-$\textbf{b}$-$\textbf{c}$ coordinate system.
\section{The non-coplanar order}
Through the above methods, we find a non-coplanar order can be stabilized by the competition between negative $K$ and small  positive $J_2$.
We take a representative point ($K=-1$, $J_2 = 1$, $h=0.05$) as an example to illustrate its typical spin configuration in Fig.\ref{fig2} (a).
The configuration has $C_3$ rotational symmetry around the $\mathbf{c}$-axis.
There are nine spins in a magnetic unit cell ($N = 9$) and the ordering wave vector is located at the $2\mathbf{M}/3$ point, as shown in Fig.\ref{fig2} (b).
Since the skyrmion crystal has an integer topological number in a magnetic unit cell,
first, we calculate the solid angle $\Omega_\Delta$ of each elementary triangle.
In the discrete lattice, it can be obtained as\cite{BERG1981}  
\begin{equation} 
\cos\left(\frac{\Omega_\Delta}{2}\right)=\frac{1+ \textbf{S}_i\cdot  \textbf{S}_j+ \textbf{S}_i\cdot  \textbf{S}_k+ \textbf{S}_j\cdot  \textbf{S}_k}{\sqrt{2\left( 1+ \textbf{S}_i \textbf{S}_j\right) \left( 1+ \textbf{S}_i \textbf{S}_k\right) \left( 1+ \textbf{S}_j \textbf{S}_k\right)}},
\end{equation} 
where $\textbf{S}_i$ is the spin at site $i$.
The sign of $\Omega_\Delta$ is determined as $\rm{sign}(\Omega_\Delta)=\rm{sign}\left[  \textbf{S}_i\cdot\left( \textbf{S}_j\times  \textbf{S}_k\right) \right] $. Note that on each triangle, the $i$, $j$, and $k$ are arranged counterclockwise. The topological number $Q$ is the sum of solid angles in a magnetic unit cell,
\begin{equation} 
Q=\frac{1}{4\pi}\sum_\Delta\Omega_\Delta.
\end{equation} 
The topological number $Q$ of the non-coplanar order can be either $\pm 1$ or $\pm 2$ (for details, see the next paragraph), and the positive and negative signs can be selected by magnetic fields of different signs. 
The magnetic field is not the cause of the formation of this magnetic order, but it can eliminate the degeneracy when all spins are reversed.
\begin{figure}[h!]	
	\begin{minipage}{1\linewidth}
		\centerline{\includegraphics[width=0.95\textwidth]{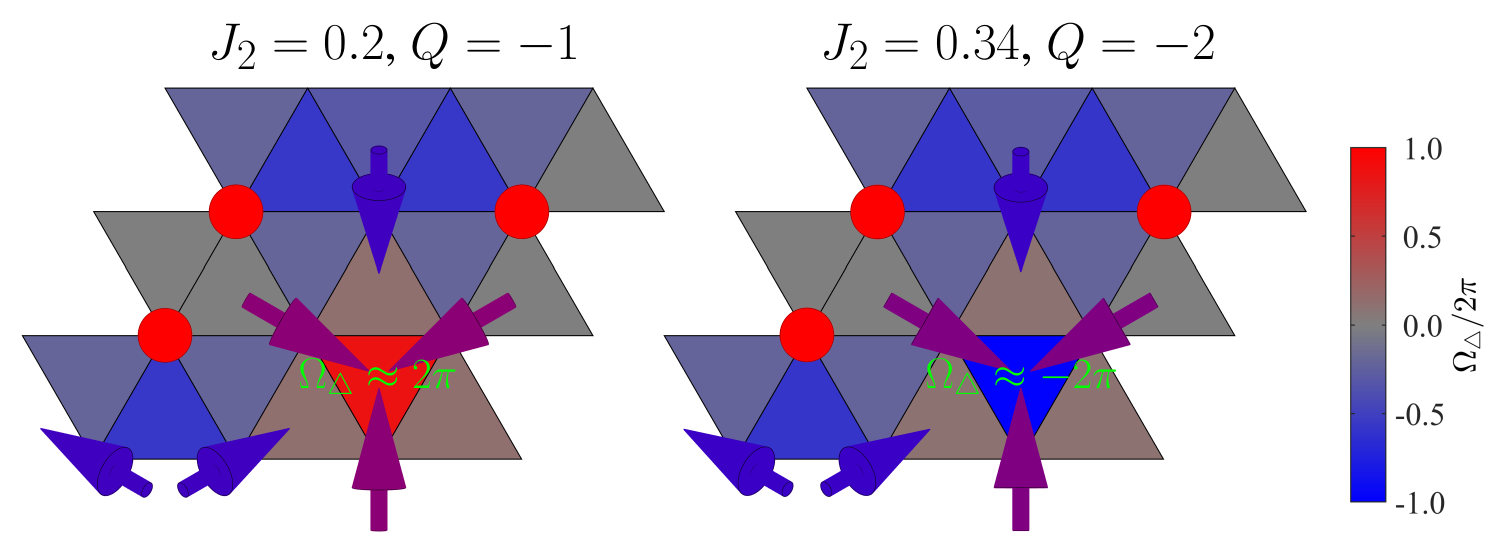}}
	\end{minipage}
	\caption{We compared the solid angle $\Omega_{\triangle}$ distribution when topological number is -1 and -2. Three spins are almost in-plane and they have the same out of plane component $S_c$. 
		As $J_2$ increases,
		the $S_c$ continuously changes from a small positive value to a small negative value, the solid angle formed by three spins 
		changes as $\left(<2\pi\right) \rightarrow 2\pi \rightarrow -2\pi  \rightarrow \left(> -2\pi\right)$.
		Thus, the topological number (in unit of $4 \pi$) as the sum of solid angle reduce by 1.
	}
	\label{fig3a}
\end{figure}

Now we illustrate the influence of $J_2$ interaction. 
The ground state of the pure Kitaev model is ferromagnetic order\cite{Becker2015}.
As shown in Fig.~\ref{fig2}(c), after fixing $K=-1$ and $h=0.05$, a small $J_2$ ($\approx 0.016$) will induce the non-coplanar order.
A narrow transition area has been identified before entering the non-coplanar order, and we omit its details.
Next, we continue to increase $J_2$. As shown in Fig.~\ref{fig2}(d), at least when $J_2$ is less than 5, the non-coplanar order has always existed as the ground state. 
Of note is that the topological number will undergo a change as $J_2$ increases.
As shown in Fig.~\ref{fig2}(e) and (f), when $J_2 \in [0.016, 0.326]$ the topological number is $-1$, and it becomes $-2$ when  $J_2$ is larger than 0.326.
Since the energy and its derivatives are continuous at $J_2 \approx 0.326$, to understand what changes have occurred, we compared the solid angle distribution before and after such point. 
As shown in Fig.~\ref{fig3a},
the jump of topological numbers is due to the continuous variation of three  almost in-plane spins.
They have the same out of plane components $S_c$, and as $S_c$ continuously changes from a small positive value to a small negative value, the solid angle formed by three spins will decrease by $4\pi$.

\begin{figure}[h!]	
	\begin{minipage}{1\linewidth}
		\centerline{\includegraphics[width=0.8\textwidth]{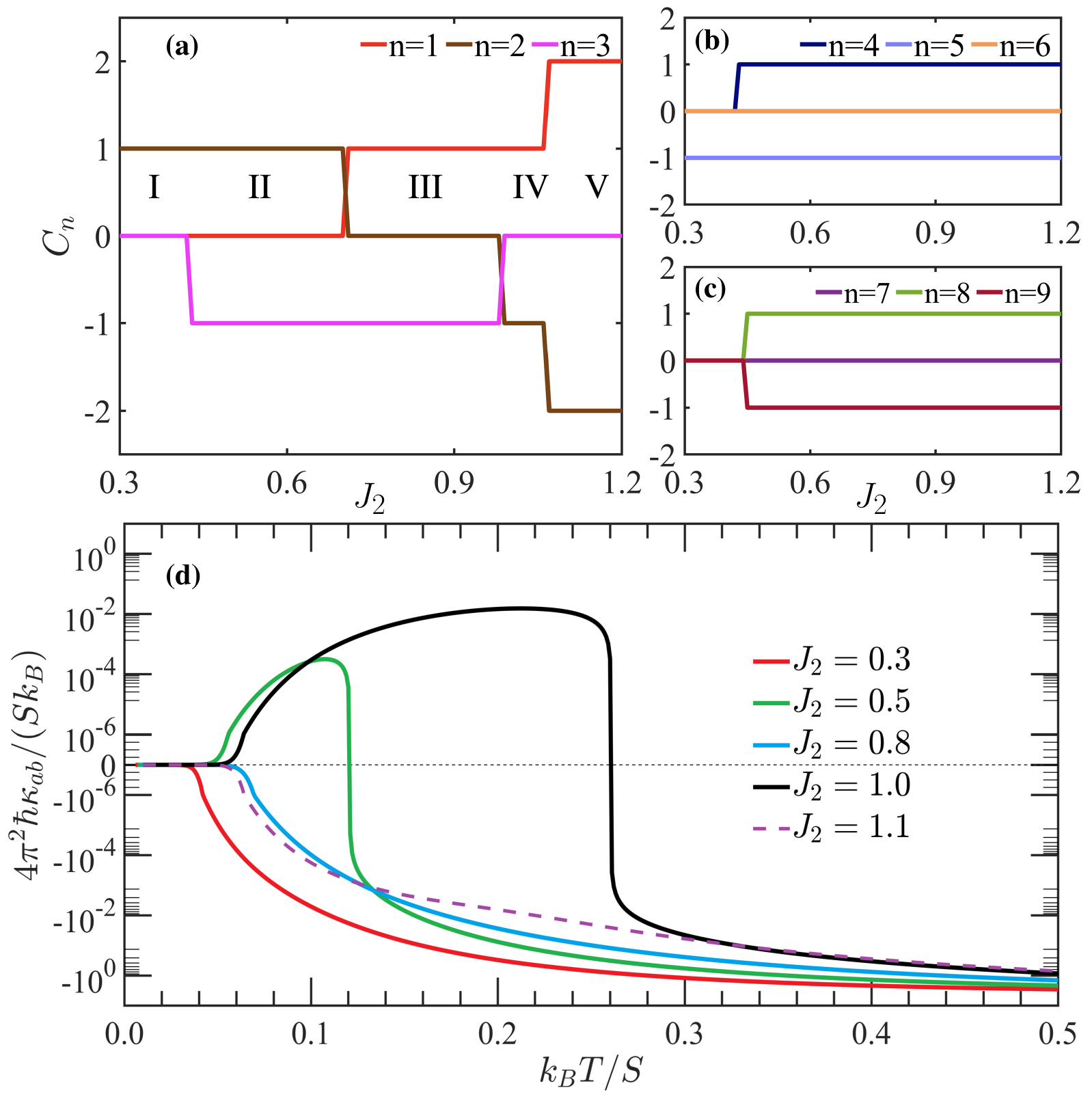}}
	\end{minipage}
	\caption{(a)-(c) Chern numbers as a function of $J_2$, and $C_1,C_2,...,C_9 $ correspond with the lowest to the highest band, respectively. 
		Five areas are distinguished by different Chern numbers. We marked them with Roman numerals \MakeUppercase{\romannumeral1}-\MakeUppercase{\romannumeral5} in (a).
		(d) The magnon thermal Hall conductivity as a function of temperature for variant $J_2$, which belong to different areas in (a). }
	\label{fig4}
\end{figure}
\section{Topological magnon and thermal Hall effect}
To consider magnon excitations in our non-coplanar order, we use the linear spin-wave theory. Its details can be found in Appendix.~\ref{appendix}.
If a magnon band is separate from the upper and lower energy bands, we can capture its topological properties by calculating the Chern number.
The Chern number of the band $n$ ($C_n$) is defined as the integral of the Berry curvature $\Omega_{n,\textbf{k}}$ over the first Brillouin zone (FBZ) of a magnetic unit cell 
\begin{equation}
C_n=\frac{1}{2\pi}\int_{\rm FBZ}\Omega_{n,\textbf{k}}d^2\textbf{k}.
\end{equation}
The Berry curvature can be calculated as\cite{Mook2019}
\begin{equation}
\Omega_{n,\mathbf{k}}=-2\operatorname{Im}\sum_{\overset{m=1}{m\neq n}}^{2N}\frac{\left(GT_\mathbf{k}^\dagger\partial_a \mathcal{H}_{\mathbf{k}}T_{\mathbf{k}}\right)_{nm}\left(GT_\mathbf{k}^\dagger\partial_b\mathcal{H}_{\mathbf{k}}T_{\mathbf{k}}\right)_{mn}}{\left[\left(G \mathcal{E}_{\mathbf{k}}\right)_{nn}-\left(G\mathcal{E}_{\mathbf{k}}\right)_{mm}\right]^2},
\end{equation}
where $G$, $\mathcal{H}_{\mathbf{k}}$, $T_{\mathbf{k}}$ and $\mathcal{E}_{\mathbf{k}}$ are all $2N \times 2N$ matrices, and $G$ is a diagonal matrix whose first $N$ diagonal elements are 1, and the final $N$ diagonal elements are
$-1$. 
The definitions of the rest matrices are shown in Appendix.~\ref{appendix}. 

In this section, we mainly focus on the impact of $J_2$, and consider a moderate parameter range $J_2 \in \left[0.3 ,1.2\right]$ since there are degeneracy points in the energy band when $J_2$ is smaller.
Fig.~\ref{fig4}(a)-(c), show the variation of Chern number with $J_2$ while fixing $K = -1$, $h = 0.05$. 
The nonzero Chern numbers are widely present within the parameter range, indicating the existence of non-trivial band topology.
The Chern number of the lowest three bands $(C_1,C_2,C_3)$ changes with the increase of $J_2$  
as $(0,1,0)\rightarrow(0,1,-1)\rightarrow(1,0,-1)\rightarrow(1,-1,0)\rightarrow(2,-2,0)$, sequentially.
Five different areas are distinguished, and we marked them with Roman numerals \MakeUppercase{\romannumeral1}-\MakeUppercase{\romannumeral5} in Fig.~\ref{fig4}(a).
Specifically, the real space topological number $Q$ changes from $-1$ to $-2$ when $J_2 \approx 0.326$, but there is no corresponding topological phase transition occurring at this point.  
Topological phase transitions mostly occur on the three lowest energy bands, for the rest bands, 
only $C_4$, $C_8$ and $C_9$ show jumps at $J_2 \approx 0.43$.
Since magnons follow the Bose-Einstein distribution, they prefer to stay in low energy states at low temperatures. The above fact highlights the importance of the role played by the lower energy bands.
\begin{figure}[h!]	
	\begin{minipage}{1\linewidth}
		\centerline{\includegraphics[width=0.95\textwidth]{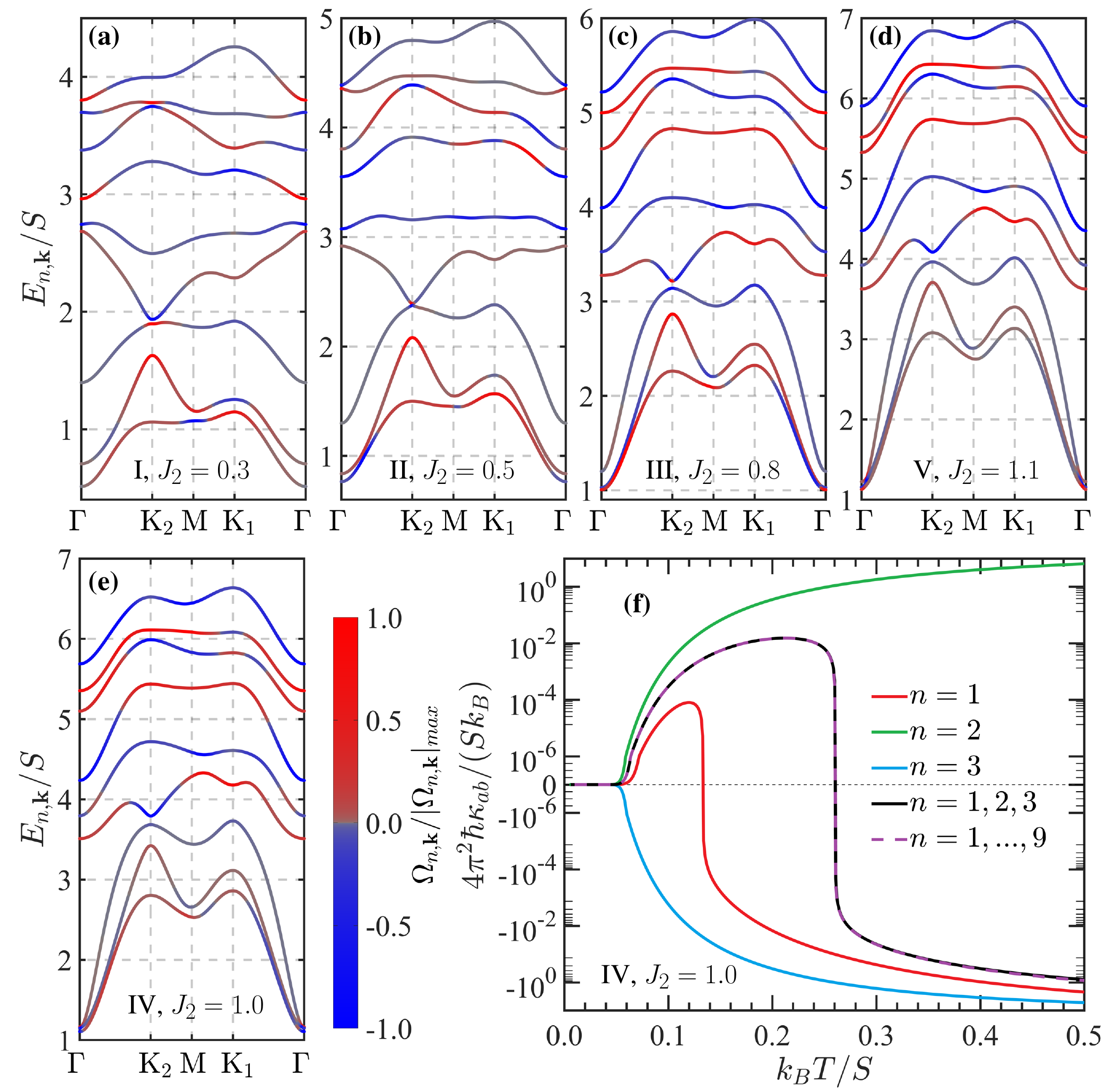}}
	\end{minipage}
	\caption{(a-e) Dispersion relations of different values of $J_2$ when $K= -1$, $h=0.05$. 
		They belong to different areas in Fig.~\ref{fig4}(a) respectively. 
		The line color stands for the normalized Berry curvature of each band. (f) The magnon thermal Hall conductivity $\kappa_{ab}$  as a function of $k_BT$ at the same parameter point with (e). 
		The curves $n = 1$, $n = 2$, and $n = 3$ are the results coming from the single band $n$, and 
		$n = 1 ,2,3$ is the sum of them. $n = 1 ,\cdots, 9$ is the overall result which includes nine bands.
	}
	\label{fig5}
\end{figure}

The thermal Hall conductivity (THC) as an observable physical quantity in experiments is related to the magnon band topology \cite{Mook2022}. 
The THC can be obtained as follows with the help of Berry curvature\cite{Matsumoto2014},
\begin{equation} \label{eq7}
\kappa_{ab}=-\frac{k_B^2T}{(2\pi)^2 \hbar}\sum_{n=1}^{N}\int_{\textbf{k}\in \rm FBZ}\left\lbrace c_2  \left[\rho\left(E_{n,\textbf{k}}\right)\right]-\frac{\pi^2}{3}\right\rbrace \Omega_{n,\textbf{k}}d^2 \mathbf{k},
\end{equation}
where $T$ is temperature, $N$ is the number of sub-lattices, and $\rho\left(E_{n,\textbf{k}}\right)$ is the Bose distribution function $\rho(E_{n,\textbf{k}})= \left( e^{E_{n,\textbf{k}}/k_BT}-1\right) ^{-1}$.
The weighting function
 $c_2(x)$ is defined as $c_2(x)=(1+x)\ln^2\frac{1+x}{x}-\ln^2x-2\rm{Li}_2(-x)$ with the Spence function $\rm{Li}_2(z)=-\int_{0}^{z}\ln(1-t)/t dt $.

Fig.~\ref{fig4}(d) shows temperature-dependent THC curves of representative points in different areas.
At higher temperatures ($T>0.34$), THC monotonically decreases with increasing $J_2$.
This relationship no longer exists at low temperatures.
Specifically, the curves of $J_2=0.5$ (which belongs to area \MakeUppercase{\romannumeral2}) and $J_2=1.0$ (which belongs to area \MakeUppercase{\romannumeral4}) both show a sign change as the temperature changes.
Among them, the THC of $J_2=1.0$ has a larger positive area at low temperatures.
The curves of the rest parameters maintain negative values at low temperatures, and the absolute value of $J_2=0.3$ (which belongs to area \MakeUppercase{\romannumeral1}) is much larger than that of $J_2=0.8$ (which belongs to area \MakeUppercase{\romannumeral3}) and $J_2 = 1.1$ (which belongs to area \MakeUppercase{\romannumeral5}).

To further clarify the differences in THC, we present the energy bands with corresponding Berry curvatures of each parameter point in Fig.~\ref{fig5}(a)-(e).
For all points, the lowest energy position is at the $\Gamma$ point.
The energy gap at the $\Gamma$ point indicates the lack of continuous symmetry, and the gap decreases with increasing $J_2$.  
Strictly speaking, all energy bands contribute to THC at finite temperature, as shown in Eq.~\ref{eq7}. 
However, as the $c_2$ function drops quickly \cite{Mook2014}, the THC at low temperature is most related to the Berry curvature in the lowest energy regions.
Fig.~\ref{fig5}(a) shows the result when $J_2 = 0.3$. Due to the fact that the lowest two bands both have positive Berry curvatures near the $\Gamma$ point, 
the THC has the maximum negative value at low temperatures.
As shown in Fig.~\ref{fig5}(b), although the lowest two bands have inverse Berry curvatures near the $\Gamma$ point when $J_2 = 0.5$, 
the Berry curvatures of the lowest band make greater contributions, resulting in a totally positive THC at low temperatures.
Fig.~\ref{fig5}(c) and (d) show the results of $J_2 = 0.8$ and $J_2 = 1.1$ respectively.
The contribution of Berry curvature to THC near the $\Gamma$ point is almost completely canceled out, since the Berry curvatures has opposite signs and the energetic separation of the lowest two bands is very small.

As shown in Fig.~\ref{fig5}(e), when $J_2 = 1.0$, 
the position of the band with the third-lowest energy at the $\Gamma$ point is close to the positions of bands with lower energy.
To further illustrate the positive region of THC, we consider the contributions of each of the three lowest energy bands to THC separately.
It can be seen in Fig.~\ref{fig5}(f), the THC of the lowest band $n=1$ changes sign at $k_{B}T/S \approx 0.14$. This is because the sign of Berry curvature quickly changes from negative to positive outside the $\Gamma$ point.
The second band $n=2$ consistently contributes positively to THC, and the contribution of the third band $n=3$ to THC is always negative and cannot be neglected at low temperatures.
In short, the total negative THC region stems from the combined effect of three bands.
Moreover, we calculated the sum of the contributions from the lowest three bands and compared it with the total THC, and found that there is relative consistency between the two curves at low temperatures.
	
\section{Summary}	
In summary, we found a peculiar non-coplanar order formed by the competition between bond-dependent Kitaev interaction $K$ and next-nearest neighbor Heisenberg coupling $J_2$ in the triangular lattice.
It can be seen as a miniature version of (high-Q) skyrmion crystal since its magnetic unit cell includes nine spins and has $\pm 1$($\pm 2$) topological number.
Through the linear spin-wave theory,  we studied the magnon excitations in such order.
The dispersions and the corresponding Chern numbers are obtained as well.
Multiple topological phases (areas) are distinguished by the Chern numbers.
Of note is that the change in the relative size of $J_2$ and $K$ will produce topological phase transitions although the topological number in real space remains unchanged.
Then we calculated the  thermal Hall conductivity and discussed its differences in different areas.
We found that  the  thermal Hall conductivity is dominated by the Berry curvature in the lowest bands at low temperatures, and in certain regions, the signs of thermal Hall conductivity will change with topological phase transitions or temperature changes.
Based on these results, we hope our work will enlighten the future research of magnonics on trigonal Kitaev materials.

\section*{Acknowledgments}
We thank Xuan-yu Wang, Rui-bo Wang and Jize Zhao for their useful feedback on the
manuscript. 
This research was supported in part by Supercomputing Center of Lanzhou University.

\appendix
\section{Linear spin wave theory}\label{appendix}
When considering the magnon excitations in a non-coplanar configuration with multiple spins in a magnetic unit cell, we perform the Holstein-Primakoff (HP) transformation with 
the quantization axis based on spin orientation. 
In our global coordinate system, the orientation of spin $i$ is
\begin{equation}
\mathbf{S}_i = \left(S_{i}^{a},S_{i}^{b},S_{i}^{c}\right) =  \left(\sin\theta_{i}\cos\phi_{i},\sin\theta_{i}\sin\phi_{i},\cos\theta_{i}\right). 
\end{equation}
With the help of polar angles, we can find a rotation matrix
\begin{equation}
\begin{aligned}
\label{rotationmatrix}
R_{i}=\left[\begin{array}{ccc}\cos\theta_{i}\cos\phi_{i} & -\sin\phi_{i} & \sin\theta_{i}\cos\phi_{i}\\
\cos\theta_{i}\sin\phi_{i} & \cos\phi_{i} & \sin\theta_{i}\sin\phi_{i}\\
-\sin\theta_{i} & 0 & \cos\theta_{i}
\end{array}\right]
\end{aligned},
\end{equation}
and determine a local coordinate system $\tilde{\mathbf{a}}$-${\tilde{\mathbf{ b}}}$-${\tilde{\mathbf{c}}}$ that satisfying $ \left(S_{i}^{{a}},S_{i}^{{b}},S_{i}^{{c}}\right)^{T} = R_{i}  \left(S_{i}^{\tilde{a}},S_{i}^{\tilde{b}},S_{i}^{\tilde{c}}\right)^{T}$. 
The spin model Eq.~[\ref{eq1:H}]  can be rewritten as 
\begin{eqnarray}\label{eqB5}
\mathcal{H} &=  \sum_{\langle i,j \rangle,\langle \langle i,j \rangle \rangle}{\mathbf{S}}_{i}^{T}  {\cdot}  { J}_{ij}  {\cdot} {\mathbf{S}}_{j} 
-\sum_{i} \mathbf{h} {\cdot}  {\mathbf{S}}_i,
\end{eqnarray}
where $J_2$ and $K$ terms are uniformly written into the interaction matrix ${ J}_{ij} $. 
After performing coordinate transformation on all spins separately, the model has the following form \cite{Zhang2021}
\begin{eqnarray}\label{Eq4}
\mathcal{H} &=  \sum_{\langle i,j \rangle,\langle \langle i,j \rangle \rangle} \tilde{\mathbf{S}}_{i}^{T} R_{i}^{T} {\cdot}  R_{i}{\tilde J}_{ij} R_{j}^{T} {\cdot} R_{j}\tilde{\mathbf{S}}_{j} 
-\sum_{i} \tilde{\mathbf{h}} R_{i}^{T} {\cdot}  R_{i} \tilde{\mathbf{S}}_i.
\end{eqnarray}
The Holstein-Primakoff expansion\cite{Holstein1940} on spin $i$ is
\begin{equation}\begin{aligned}
&S_{i}^{\tilde{c}} =S-b_{i}^{\dagger}b_{i}=S-n_{i}  \\
&S_{i}^{\tilde{a}} =\frac{\sqrt{2S-n_{i}}b_{i}+b_{i}^{\dagger}\sqrt{2S-n_{i}}}{2}\approx\sqrt{\frac{S}{2}}\left(b_{i}+b_{i}^{\dagger}\right)  \\
&S_{i}^{\tilde{b}} =\frac{\sqrt{2S-n_{i}}b_{i}-b_{i}^{\dagger}\sqrt{2S-n_{i}}}{2}\approx-i\sqrt{\frac{S}{2}}\left(b_{i}-b_{i}^{\dagger}\right),
\end{aligned}\end{equation}
and we keep only the lowest
order of the boson operator.
Then we substitute $\tilde{\mathbf{S}}_{i}=\left(S_{i}^{\tilde{a}},S_{i}^{\tilde{b}},S_{i}^{\tilde{c}}\right)^{T}$  for each spin in a magnetic unit cell and 
apply the Fourier transformation 
\begin{equation}
b_{i} = b_{n,j}=\frac{1}{\sqrt{L}}\sum_{\mathbf{k} \in \rm FBZ}b_{j}(k)e^{i\mathbf{k}\cdot(\mathbf{V}_{n}+\mathbf{r}_{j})},
\end{equation}
where $n$ is index of unit cell, $j$ ($\in 1,\cdots N$) marks the sub-lattice inside unit cell, $L$ is the total number of unit cells and $\mathbf{V}_n$ is Bravais lattice coordinate. 
Finally, we can obtain the spin wave Hamiltonian, and we focus on the quadratic term,
\begin{equation}
\mathcal{H}_2=\frac 12\sum_\mathbf{k}\psi_\mathbf{k}^\dagger\mathcal{H}_{\mathbf{k}}\psi_\mathbf{k},
\end{equation}
where $\psi_{\mathbf{k}}^{\dagger}=\left(a_{1,\mathbf{k}}^{\dagger},a_{2,\mathbf{k}}^{\dagger},\cdots,a_{N,\mathbf{k}}^{\dagger},a_{1,-\mathbf{k}},a_{2,-k},\cdots,a_{N,-\mathbf{k}}\right)$ and  $\mathcal{H}_\mathbf{k}$ is a 2\textit{N}$\times$2\textit{N} matrix.
We can use a transformation matrix $T_\mathbf{k}$ to diagonalize the matrix $\mathcal{H}_\mathbf{k}$ \cite{COLPA1978327},
\begin{equation}\mathcal{E}_{\mathbf{k}}={T_\mathbf{k}^{\dagger}\mathcal{H}_{\mathbf{k}}T_{\mathbf{k}}},\end{equation}
where $\mathcal{E}_{\mathbf{k}} = diag \left( E _{1,\textbf{k}}, E _{2,\textbf{k}},\cdots,E _{N,\textbf{k}},E _{1,-\textbf{k}},E _{2,-\textbf{k}},\cdots,E _{N,-\textbf{k}}  \right) $ contains the magnon dispersions.

\bibliography{iopart-num}

\end{document}